# Quantification of entanglement entropy in helium by the Schmidt-Slater decomposition method

Chien-Hao Lin[1,2] and Yew Kam Ho[2]

[1] Department of Physics, National Taiwan University, Taipei, Taiwan
[2] Institute of Atomic and Molecular Sciences, Academia Sinica, Taipei, Taiwan

Abstract

In this work, we present an investigation on the spatial entanglement entropies in the helium atom by using highly correlated Hylleraas functions to represent the $S$-wave states. Singlet-spin $1sns\ ^1S^e$ states (with $n = 1$ to 6) and triplet-spin $1sns\ ^3S^e$ states (with $n = 2$ to 6) are investigated. As a measure on the spatial entanglement, von Neumann entropy and linear entropy are calculated. Furthermore, we apply the Schmidt-Slater decomposition method on the two-electron wave functions, and obtain eigenvalues of the one-particle reduced density matrix, from which the linear entropy and von Neumann entropy can be determined.

e-mail address: ykho@pub.iams.sinica.edu.tw

## I. Introduction

Entanglement measures in atomic systems have attracted considerable attention in recent years [1-9]. Related investigations on interacting bosons [10-15] and on quantum dots [16, 17] have also been reported in the literature. Such investigations help us to shed light on the relationship between entanglement and correlation effects. Studies of quantum entanglement also play an important role in quantum cryptography [18], quantum teleportation [19], quantum information and quantum computation [20]. Tichy *et al.* [21] reviewed the recent developments in entanglement for atomic and molecular systems. In the present work, we emphasize on calculations of entanglement entropies, such as linear entropy and von Neumann entropy, in the helium atom, as they are practical and quantitative measures for the amount of entanglement in two-electron systems. Highly-correlated Hylleraas-type wave functions are utilized to represent the state wave functions, and take into account of the correlation effects. The inter-electronic terms in the Hylleraas-type basis greatly improve the quality of wave functions. To calculate the entropies, we carry out partial wave expansion on the Hylleraas basis, and apply the Schmidt-Slater decomposition method on each of the partial wave functions. Then the eigenvalues of the reduced density matrix can be obtained, and from which the entropies can be determined. This method is similar to that by Kościk [15], but we have improved the mythology such that it is more computationally efficient and with accurate results. In our present work we have calculated the entropies for some *S*-wave 1*sns* singlet-spin states (with $n = 1$ to 6), and 1*sns* triplet-spin states (with $n = 2$ to 6) of the helium atom. Atomic units (a.u.) are used throughout the present work. Computations are carried out, for the most part, with quadruple precision. For some critical cases, calculations are performed with multiple precision algorithm, with up to about 100 digits in the word length are used.

## II. Theoretical Method

The non-relativistic Hamiltonian (in atomic units) describing the helium atom is

$$H = -\frac{1}{2}\nabla_1^2 - \frac{1}{2}\nabla_2^2 - \frac{Z}{r_1} - \frac{Z}{r_2} + \frac{1}{r_{12}} \quad , \qquad (1)$$

where 1 and 2 denote the two electron 1 and 2, respectively, and $r_{ij}$ is the relative distance between the particle *i* and *j*. The Hamiltonian with $Z = 2$ is for the helium atom. For *S*-states we use Hylleraas-type wave functions to describe the system,

$$\Psi_{kmn} = \sum_{kmn} C_{kmn} \left( \exp[-\alpha r_1 - \beta r_2] r_{12}^k r_1^m r_2^n \pm (1 \leftrightarrow 2) \right), \qquad (2)$$

with $k+m+n \leq \omega$, and $\omega$, $k$, $m$ and $n$ are positive integers or zero. In the present work we use $\omega$ up to 15, implying the total number of terms in Eq. (2) is $N_t = 444$ for the singlet-spin states and $\omega = 16$, $N_t = 444$ for triplet-spin states, respectively.

The quantum entanglement of an atomic system can be quantified with entropies, such as von Neumann entropy and linear entropy. The von Neumann entropy of the spatial entanglement for a two-electron system has the form (see [3, 4, 6, 7, 9] for example).

$$S_{vN} = -\text{Tr}(\rho_{red} \log_2 \rho_{red}) \ , \quad (3)$$

where $\rho_{red}$ is the one-particle reduced density matrix, and Tr has the implication of tracing out the degrees of freedom of one of the two electrons in the system. In addition, the linear entropy is defined as

$$S_L = 1 - \text{Tr}\rho_{red}^2 \ . \quad (4)$$

Once the eigenvalues $\Lambda$ of the reduced density matrix are solved, one can directly acquire the entropies by using the equations above (more details will be discussed later). To calculate eigenvalues of the reduced density matrix, we adopt the Schmidt-Slater decomposition method. The two-electron wave function now is a function of $\mathbf{r}_1$ and $\mathbf{r}_2$, and typically it can be decomposed into a sum of products by partial wave expansion, as

$$\Psi(\mathbf{r}_1, \mathbf{r}_2) = \sum_{i=0}^{\infty} a_i \phi_i(\mathbf{r}_1) \phi_i(\mathbf{r}_2) \ . \quad (5)$$

In addition, for a spherically symmetric system, the wave function above can be expanded further in a series of Legendre polynomials. As a result, the angular correlation of this spherical system is represented by the Legendre polynomials and the coefficients $F_l$ of such expansion are now functions of scalar $r_1$ and $r_2$, as

$$\Psi(\mathbf{r}_1, \mathbf{r}_2) = \sum_{l=0}^{\infty} F_l(r_1, r_2) P_l(\cos\theta) = \sum_{l=0}^{\infty} \frac{f_l(r_1, r_2)}{r_1 r_2} P_l(\cos\theta) \ . \quad (6)$$

For practical computational purposes, the use of Eq. (6) is quite crucial to transform the two-electron wave function into a series of scalar functions. Also in Eq. (6) the infinite sum in $l$ must be truncated into a finite sum, and we use maximum value of 40, ie $l_{max} = 40$, in our present work. Later in the text, we will show examples for convergence when different values of $l_{max}$ are used. For a given $l$, with the help of Schmidt decomposition (for singlet-spin states) or Slater decomposition (for triplet-spin states), the function $f_l(r_1, r_2)$ can be decomposed as a sum of products of one-particle wave functions. For a real and symmetric wave function (the singlet-spin cases), the function $f_l^+$ has the Schmidt decomposition:

$$f_l^+(r_1, r_2) = \sum_{n=0}^{\infty} \lambda_{nl}^+ u_{nl}^+(r_1) u_{nl}^+(r_2) , \qquad (7)$$

where the $u_{nl}^+$ is a set of orthonormal basis. Otherwise, if the wave function is real but antisymmetric (the triplet-spin cases), the function $f_l^-$ follows the Slater decomposition procedure:

$$f_l^-(r_1, r_2) = \sum_{n=0}^{\infty} \lambda_{nl}^- \left( u_{2n+1,l}^-(r_1) u_{2n,l}^-(r_2) - u_{2n+1,l}^-(r_2) u_{2n,l}^-(r_1) \right) , \qquad (8)$$

where $u_{nl}^-$ are also orthonormal functions. Since the Schmidt-Slater decomposition can be expressed as eigenvalue problems, for real and symmetric wave functions the eigenvalues can be determined via an integral equation

$$\int_0^{\infty} f_l^+(r_1, r_2) u_{nl}^+(r_2) dr_2 = \lambda_{nl}^+ u_{nl}^+(r_1) . \qquad (9)$$

Up to this point, our treatment is very similar to that in Ref. [15]. In order to solve the above eigenvalue problem in a form of integral equation, the author in Ref. [15] used numerical quadratures to discretize the variables $r_1$ and $r_2$ with equal subintervals, and turned the above integral equation into an algebraic eigenvalue problem. In our present work, for better efficiency and better accuracy, we first obtain the matrix elements for the reduced density matrix by projecting the functions $f_l$ (see Eq (7) or Eq (8)) onto one-particle orthonormal Laquerre polynomials. By using such Laquerre polynomials with dimension $LA_{max}$ as basis, the matrix elements $\langle L_i(r_1) | f_l(r_1, r_2) | L_j(r_2) \rangle$ can be calculated analytically in such a way that they are suitable for machine computation, whereas in Ref. [15] some accuracy may be lost due to numerical integration using quadratures. Furthermore, our analytical treatment on matrix elements for excited states is very straightforward, but using numerical quadratures for excited states, in order to take into account of the diffuseness of wave functions, a longer-range in the coordinate space must be used, and hereby it would need more quadratures points, resulting in more difficulty in numerical integrations. In our work, once the elements of the density matrix are calculated, eigenvalues can then be obtained by diagonalization of the reduced density matrix. As for the anti-symmetric cases with Slater decomposition, it is possible to transform a $2n*2n$ real anti-symmetric matrix to a block diagonal form

$$\begin{bmatrix} 0 & \lambda_1 & & 0 & \cdots & & 0 \\ -\lambda_1 & 0 & & & & & \\ & & 0 & \lambda_2 & & & 0 \\ 0 & & -\lambda_2 & 0 & & & \\ \vdots & & & & \ddots & & \vdots \\ 0 & & 0 & & \cdots & 0 & \lambda_n \\ & & & & & -\lambda_n & 0 \end{bmatrix}, \quad (10)$$

leading to eigenvalues of $\pm i\lambda_k$ for the function $f_l^-$. The size of basis set can be chosen depending on the required precision. In actual calculations, the eigenvalues $\Lambda_{nl}^\pm$ of the reduced density matrix (see Eqs. [3] and [4]) are related to $\lambda_{nl}^\pm$ in Eqs. [7] and [8]. The reduced density matrix can be expanded as

$$\rho_{red}(\mathbf{r},\mathbf{r}') = \int [\Psi(\mathbf{r},\mathbf{r}_1)]^* \Psi(\mathbf{r}_1,\mathbf{r}') d\mathbf{r}_1 \quad , \quad (11)$$

$$\rho_{red}(\mathbf{r},\mathbf{r}') = \int \sum_{l_1,l_2} \frac{f_{l_1}(r,r_1)}{rr_1} \frac{f_{l_2}(r',r_1)}{r'r_1} P_{l_1}(\cos\theta) P_{l_2}(\cos\theta') d\mathbf{r}_1. \quad (12)$$

This equation can still be expanded further with spherical harmonic functions. Let's take the symmetric case for example, we have

$$\rho_{red}(\mathbf{r},\mathbf{r}') = \int \sum_{\substack{l_1,l_2 \\ n_1,n_2}} \left(\frac{4\pi}{2l_1+1}\right)\left(\frac{4\pi}{2l_2+1}\right) \frac{\lambda_{n_1 l_1}^+ \lambda_{n_2 l_2}^+ u_{n_1 l_1}^+(r) u_{n_1 l_1}^+(r_1) u_{n_2 l_2}^+(r') u_{n_2 l_2}^+(r_1)}{rr'}$$
$$\times Y_{l_1 m_1}(\theta,\varphi) Y_{l_1 m_1}(\theta_1,\varphi_1) Y_{l_2 m_2}(\theta',\varphi') Y_{l_2 m_2}(\theta_1,\varphi_1) dr_1 d\Omega_1 \quad , \quad (13)$$

$$\rho_{red}(\mathbf{r},\mathbf{r}') = \sum_{n,l,m} \left(\frac{4\pi}{2l+1}\right)^2 \frac{(\lambda_{nl}^+)^2}{rr'} u_{nl}^+(r) u_{nl}^+(r') Y_{lm}(\theta,\varphi) Y_{lm}(\theta',\varphi') = \sum_{n,l,m} \left(\frac{4\pi \lambda_{nl}^+}{2l+1}\right)^2 \phi(\mathbf{r})\phi(\mathbf{r}'). \quad (14)$$

Thus, the relation between $\Lambda_{nl}^\pm$ and $\lambda_{nl}^\pm$ is given by

$$\Lambda_{nl}^\pm = \left(\frac{4\pi \lambda_{nl}^\pm}{2l+1}\right)^2. \quad (15)$$

This can also be applied to antisymmetric cases. One thing to be noticed is that the 'm' value in this case is an independent quantum number and that our system is with (2l+1)-fold degenerate. The von Neumann entropy for spatial entanglement is then given by

$$S_{vN} = -\sum_{nl}(2l+1)\Lambda_{nl} \log_2 \Lambda_{nl} \quad , \quad (16)$$

and the linear entropy for spatial entanglement by

$$S_L = 1 - \sum_{nl}(2l+1)\Lambda_{nl}^2 \quad . \qquad (17)$$

Finally, we should mention that here we emphasis on the spatial (electron-electron orbital) entanglement of the two electrons in the helium atom. For entanglement due to the spin part, readers are referred to some earlier well-discussed publications [3, 14, 15].

III. **Calculations and Results**

In the present work, we calculate the von Neumann entropy and the linear entropy of the singlet-spin states from $1s^2\,^1S$ to $1s6s\,^1S$ and triplet-spin states from $1s2s\,^3S$ to $1s6s\,^3S$. Table I shows the convergence of energies and entropies in terms of the sizes of the basis sets for the $1s^2\,^1S^e$ state of the helium atom. Here, we use Hylleraas-type wave functions up to $N_t = 444$ terms ($\omega = 15$ and 16 for singlet and triplet, respectively). The maximum size of the Laquerre polynomials, denoted as $LA_{max}$, is set for 50 and the maximum size for Legendre expansion $l_{max}$ is set for 40, meaning that partial waves with $l = 0$ and up to $l = 40$ are used for the expansion in Eq. (6). Here we also use the sum of eigenvalues as a criterion for convergence. It is seen from Table I that the computed sum is very close to the theoretical value of 1.0, implying that our results have achieved great convergence and accuracy. Table II shows similar convergence for the $1s2s\,^1S^e$ state. In Table III we show the convergence of energies and entropies of the $1s2s\,^3S^e$ state. We use wave functions up to $N_t = 444$ terms ($\omega = 16$), $LA_{max} = 50$ and $l_{max} = 40$. Again, our results show that good convergence in terms of the sizes of the basis sets can be achieved for triplet-spin states. In Tables IV and V, we show convergence for the sum of eigenvalues and entropies of the $1s^2\,^1S^e$ state and $1s2s\,^3S^e$ state, respectively, in terms of different $l_{max}$ values, with $l_{max} = 5, 10, 15, 20, 25, 30, 35$ and 40. Here we keep $N_t = 444$ terms and $LA_{max} = 50$. Again, results show good convergence in terms of different maximum values used in partial wave expansion. As for convergence in terms of difference sizes of the Laquerre polynomials used in calculations for the elements in the partial wave reduced density matrix, we show in Tables VI and VII for entropies of the $1s^2\,^1S^e$ state and $1s2s\,^3S^e$ state, respectively. Here we keep $N_t = 444$ terms and $l_{max} = 40$. By examining results for $LA_{max} = 20, 25, 30, 35, 40, 45$ and 50, it is concluded that our results show excellent convergence.

In Table VIII, we summarize our results for the singlet-spin states and show a comparison of the linear entropy $S_L$ with earlier calculations [3, 4, 6, 7, 8, 15 and 22]. For the ground $1s^2\,^1S^e$ state, in addition to our present result, we also show the earlier results that were obtained using Eq. (4) and with which four-electron integrals were used [3, 4, 8]. Table VIII also shows results that were obtained by employing configuration interaction wave functions with B-splines basis [6], with products of Slater-type orbitals (STO) [7], with Hylleraas type basis functions [15] and with Gaussian type basis sets [22]. It is seen that our present result of $S_L = 0.01591564$ calculated by using Eq. (17) is practically identical to that in Ref. [8] ($S_L = 0.0159156 \pm 0.0000010$) calculated by employing Hylleraas functions and with the use of Eq.(4). In the present approach, no four-electron integrals are needed for evaluation of linear entropy. In Table VIII we also show results for other excited singlet-spin $1sns\ ^1S^e$ states and excited triplet-spin $1sns\ ^3S^e$ states with $n = 2$ to 6. We also show comparisons of our results with those in earlier publications [3, 4, 6, 7, 9]. From the results shown in Table VIII, it is seen that the linear entropies for the $^1S^e$ states increase as the state energy increases (increasing $n$), but for the $^3S^e$ states the linear entropy is decreasing for increasing $n$. Both the excited singlet-spin states and triplet-spin states are approaching the saturated value of 0.5, the non-interacting limit for the two electrons. Later in the text, we will come back to discuss the implication of such finding.

In Table IX, we show our results for von Neumann entropy for the ground and excited states of the helium atom. For the ground $1s^2\ ^1S^e$ state, our result gives $S_{vN} = 0.08489987$. In Table IX we also show results by Benenti *et. al.* [7] who used CI basis for wave functions ($S_{vN} = 0.0785$) and from Ref. [9] in which CI functions with B-splines basis were used ($S_{vN} = 0.085022$). Huang *et. al.* [22] obtained $S_{vN} = 0.0675$ by using Gaussian basis with Eq. (16). Hofer [23] obtained $S_{vN} = 0.06749889$ using Gaussian type basis sets. It indicates that the value in Ref. [9] is closer to our present result than that of Ref. [7]. In Ref. [15], $S_{vN}$ for some helium states were also calculated, but they were expressed in graphic forms, and hence no direct numerical comparisons with the $S_{vN}$ results in Ref. [15] can be made here. In Table IX, we show von Neumann entropy for excited singlet-spin $1sns\ ^1S^e$ states and excited triplet-spin $1sns\ ^3S^e$ states with $n = 2$ to 6. We also show comparisons of our results with those in earlier publications [7, 9]. From these results, we observe that the von Neumann entropy for excited $^1S^e$ states increases as the state energy increases (increasing $n$), while $S_{vN}$ for the excited $^3S^e$ states decreases for increasing energy (increasing $n$). The $S_{vN}$ for both the excited singlet-spin states and triplet-spin states are approaching the saturated value of 1.0, the non-interacting limit for $S_{vN}$.

Next, from our results we examine the relationship between entanglement, entropy and correlation, and try to shed night on a dilemma presented to us for excited states in the helium atom. While for the ground state, we can relate entropy to entanglement directly, but if we relate entropy to entanglement directly for excited states, we will encounter a dilemma situation [7]. For the singlet-spin excited states (1sns), entropy increases when the energy of the excited state increases (increasing *n*). But 'intuition' tells us that 'correlation energy' would decrease when the energy of the excited state increases (the two electrons are farther apart along the Rydberg series as *n* increases). Hence 'entanglement' and 'correlation' seem go opposite ways for singlet-spin excited states if entropy is directly related to entanglement. As for triplet-spin excited states, from our present calculations and from those in Ref. [7], it is observed that both $S_L$ and $S_{vN}$ decrease as the quantum number *n* of the excited state along the Rydberg series increases. In order to have a consistent way to interpret entropy, entanglement and correlation, Benenti *et. al.* [7] proposed an alternate definition for entanglement. Entanglement is defined as "the distance between the calculated entropy (for a state with principal quantum number *n*) and the entropy without interaction." In mathematical term, entanglement ε is defined, in a form of absolute value, as

$$\varepsilon |\Phi_{AB}\rangle = |S(\rho_A) - S^{(0)}(\rho_A)| \quad , \qquad (18)$$

with $S^{(0)}(\rho_A)$ being the entropy for the non-interacting limit.

Under such an alternate definition, Benenti *et. al.* [7] concluded that entanglement would drop with energy (or would increase for decreasing energy). Their conclusion seems to be in line with our intuition as when the two electrons are separated farther apart as the principal quantum number *n* of the excited electron increases; correlation becomes weaker, leading to smaller quantification value of entanglement. For the ground state, the non-interacting von Neumann entropy and linear entropy both have zero values (the Hartree-Fock values). For excited states, the non-interacting limits for von Neumann entropy and linear entropy have values of 1.0 and 0.5, respectively. Figure 1 shows entanglement, under the alternate definition obtained using Eq. (18), *vs* various states with principal quantum number *n*, in log-log plot. It is seen that entanglement is decreasing for increasing *n*, and our finding is consistent with the conclusion made in Ref. [7].

### IV. Summary and Conclusion

In this work, we have focused on the investigation of quantum entanglement in a natural atomic system; the helium atom. Using highly correlated Hylleraas basis functions, we have obtained accurate wave functions for the ground state and some excited states by optimized their energies individually. Once the wave functions for such states are obtained, we employ them together with applying the Schmidt-Slater decomposition method for quantification of entanglement, and for measure of von Neumann entropy and linear entropy. Results are obtained for singlet-spin states from $1s^2$ $^1S$ to $1s6s$ $^1S$ and triplet-spin states from

$1s2s\ ^3S$ to $1s6s\ ^3S$. We believe our results are quite accurate that they can be treated as benchmark values for future references on entropies in the two-electron helium atom. In the future, entanglement and entropy for doubly-excited (resonance) states in two-electron atomic systems, combined with the stabilization method [24, 25], can be investigated using the computational procedures developed in the present work.

V. Acknowledgement

This work was supported by the Ministry of Science and Technology of Taiwan.

Table I. Energy and entropies of He $1s^2\ ^1S^e$ state ($LA_{max}$=50, $l_{max}$=40)

| ω | $N_t$ | Energy (a. u.) | Sum of Eigenvalues | $S_L$ | $S_{vN}$ |
|---|---|---|---|---|---|
| 5 | 34 | -2.9037212928 | 0.9999985801 | 0.0159157859 | 0.0848691324 |
| 6 | 50 | -2.9037237676 | 0.9999999907 | 0.0159165492 | 0.0849046320 |
| 7 | 70 | -2.9037241789 | 0.9999999908 | 0.0159159303 | 0.0849013555 |
| 8 | 95 | -2.9037243146 | 0.9999999924 | 0.0159157560 | 0.0849003905 |
| 9 | 125 | -2.9037243542 | 0.9999999926 | 0.0159156930 | 0.0849000513 |
| 10 | 161 | -2.9037243681 | 0.9999999934 | 0.0159156696 | 0.0848999342 |
| 11 | 203 | -2.9037243733 | 0.9999999938 | 0.0159156586 | 0.0848998832 |
| 12 | 252 | -2.9037243754 | 0.9999999945 | 0.0159156522 | 0.0848998688 |
| 13 | 308 | -2.9037243761 | 0.9999999950 | 0.0159156479 | 0.0848998614 |
| 14 | 372 | -2.9037243766 | 0.9999999954 | 0.0159156486 | 0.0848998775 |
| 15 | 444 | -2.9037243768 | 0.9999999955 | 0.0159156476 | 0.0848998788 |

Table II. Energy and entropies of He $1s2s$ $^1S^e$ state ($LA_{max}$ =50, $l_{max}$=40), in terms of different sizes of basis sets.

| $\omega$ | $N_t$ | Energy (a.u.) | Sum of eigenvalues | $S_L$ | $S_{vN}$ |
|---|---|---|---|---|---|
| 5 | 34 | -2.145941590 | 0.9999999940 | 0.4887237657 | 0.9919482366 |
| 6 | 50 | -2.145960961 | 0.9999999914 | 0.4886855171 | 0.9918459593 |
| 7 | 70 | -2.145971072 | 0.9999999944 | 0.4887372020 | 0.9919161542 |
| 8 | 95 | -2.145972218 | 0.9999999935 | 0.4887333957 | 0.9919081881 |
| 9 | 125 | -2.145973654 | 0.9999999951 | 0.4887400877 | 0.9919172010 |
| 10 | 161 | -2.145973799 | 0.9999999952 | 0.4887402340 | 0.9919170983 |
| 11 | 203 | -2.145973975 | 0.9999999958 | 0.4887404382 | 0.9919173516 |
| 12 | 252 | -2.145974006 | 0.9999999960 | 0.4887404046 | 0.9919172484 |
| 13 | 308 | -2.145974030 | 0.9999999963 | 0.4887404487 | 0.9919172938 |
| 14 | 372 | -2.145974038 | 0.9999999964 | 0.4887404079 | 0.9919172142 |
| 15 | 444 | -2.145974042 | 0.9999999967 | 0.4887404076 | 0.9919172194 |

Table III. Energy and entropies for the He $1s2s\,^3S^e$ state ($LA_{max}=50$, $l_{max}=40$), in terms of different sizes of basis sets.

| $\omega$ | $N_t$ | Energy (a.u.) | Sum of Eigenvalues | $S_L$ | $S_{vN}$ |
|---|---|---|---|---|---|
| 6 | 34 | -2.17522843950 | 0.9999999978 | 0.500376244485 | 1.0055306698 |
| 7 | 50 | -2.17522911749 | 0.9999999979 | 0.500376056795 | 1.0055282803 |
| 8 | 70 | -2.17522934285 | 0.9999999980 | 0.500375944692 | 1.0055268826 |
| 9 | 95 | -2.17522936814 | 0.9999999981 | 0.500375937542 | 1.0055268224 |
| 10 | 125 | -2.17522937531 | 0.9999999983 | 0.500375933463 | 1.0055267672 |
| 11 | 161 | -2.17522937718 | 0.9999999985 | 0.500375934042 | 1.0055267839 |
| 12 | 203 | -2.17522937789 | 0.9999999986 | 0.500375934094 | 1.0055267913 |
| 13 | 252 | -2.17522937810 | 0.9999999988 | 0.500375934087 | 1.0055267976 |
| 14 | 308 | -2.17522937818 | 0.9999999989 | 0.500375934086 | 1.0055268012 |
| 15 | 372 | -2.17522937821 | 0.9999999990 | 0.500375934077 | 1.0055268054 |
| 16 | 444 | -2.17522937822 | 0.9999999991 | 0.500375934075 | 1.0055268077 |

Table IV. Convergence for linear entropy and von Neumann entropy,
in terms of $l_{max}$ for the $1s^2\ ^1S$ state ($\omega=15$, $N_t=444$, $LA_{max}=50$).

| $l_{max}$ | Sum of eigenvalues | $S_L$ | $S_{vN}$ |
|---|---|---|---|
| 5 | 0.999998815400 | 0.0159156476327 | 0.08486755928 |
| 10 | 0.999999937686 | 0.0159156476327 | 0.08489790841 |
| 15 | 0.999999987114 | 0.0159156476327 | 0.08489955694 |
| 20 | 0.999999993592 | 0.0159156476327 | 0.08489979801 |
| 25 | 0.999999994982 | 0.0159156476327 | 0.08489985357 |
| 30 | 0.999999995378 | 0.0159156476327 | 0.08489987031 |
| 35 | 0.999999995515 | 0.0159156476327 | 0.08489987635 |
| 40 | 0.999999995569 | 0.0159156476327 | 0.08489987884 |

Table V. Convergence for linear entropy and von Neumann entropy, in terms of $l_{max}$ for the $1s2s\ ^3S^e$ state ($\omega=16$, $N_t=444$, $LA_{max}=50$).

| $l_{max}$ | Sum of eigenvalues | $S_L$ | $S_{vN}$ |
|---|---|---|---|
| 5 | 0.999999986557 | 0.500375934075 | 1.005526391041 |
| 10 | 0.999999998891 | 0.500375934075 | 1.005526800615 |
| 15 | 0.999999999053 | 0.500375934075 | 1.005526807252 |
| 20 | 0.999999999061 | 0.500375934075 | 1.005526807631 |
| 25 | 0.999999999061 | 0.500375934075 | 1.005526807646 |
| 30 | 0.999999999062 | 0.500375934075 | 1.005526807676 |
| 35 | 0.999999999062 | 0.500375934075 | 1.005526807677 |
| 40 | 0.999999999062 | 0.500375934075 | 1.005526807677 |

Table VI. Convergence for linear entropy and von Neumann entropy, in terms of $LA_{max}$ for the ground $1s^2\ {}^1S^e$ state ($\omega=15$, $N_t=444$, $l_{max}=40$).

| $LA_{max}$ | Sum of eigenvalues | $S_L$ | $S_{vN}$ |
|---|---|---|---|
| 20 | 0.999999943444 | 0.015915647632707 | 0.084898248 |
| 25 | 0.999999970286 | 0.015915647632683 | 0.084899054 |
| 30 | 0.999999982216 | 0.015915647632683 | 0.084899431 |
| 35 | 0.999999988390 | 0.015915647632683 | 0.084899632 |
| 40 | 0.999999991936 | 0.015915647632683 | 0.084899752 |
| 45 | 0.999999994131 | 0.015915647632683 | 0.084899828 |
| 50 | 0.999999995569 | 0.015915647632683 | 0.084899878 |

Table VII. Convergence for linear entropy and von Neumann entropy, in terms of $LA_{max}$ for the $1s2s\ ^3S^e$ state ($\omega$=16, $N_t$=444, $l_{max}$=40).

| $LA_{max}$ | Sum of eigenvalues | $S_L$ | $S_{vN}$ |
|---|---|---|---|
| 20 | 0.999999995007 | 0.500375934075063 | 1.0055266830 |
| 25 | 0.999999998034 | 0.500375934075023 | 1.0055267744 |
| 30 | 0.999999999062 | 0.500375934075006 | 1.0055268077 |
| 35 | 0.999999999492 | 0.500375934075005 | 1.0055268223 |
| 40 | 0.999999999699 | 0.500375934075005 | 1.0055268296 |
| 45 | 0.999999999809 | 0.500375934075005 | 1.0055268337 |
| 50 | 0.999999999873 | 0.500375934075005 | 1.0055268360 |

Table VIII. Energies and linear entropies of the singlet and triplet $S$ states in the helium atom.

| State | Energy (a.u.) | | | $S_L$ | | | |
|---|---|---|---|---|---|---|---|
| | Present work | Lin et al. [6] | Dehesa et al. [3, 4] | Present work | Others | Dehesa et al.[3,4] | Benenti et. al. [7] |
| $1s^2\,^1S$ | -2.90372437 | −2.9035869 | −2.903724377 | 0.01591564 | 0.015943[a]<br>0.0159156 ± 0.0000010[b]<br>0.0159172[c]<br>0.01595052[d] | 0.015914 ±0.000044 | 0.01606 |
| $1s2s^1S$ | -2.14597404 | −2.1459653 | −2.145974046 | 0.48874040 | 0.488736[a]<br>0.488737[e] | 0.48866±0.00030 | 0.48871 |
| $1s3s^1S$ | -2.06127196 | −2.0612695 | −2.061271954 | 0.49725195 | 0.497251[a] | 0.49857±0.00097 | 0.49724 |
| $1s4s^1S$ | -2.03358497 | −2.0335856 | −2.033586653 | 0.49892499 | 0.498925[a] | 0.49892±0.00052 | 0.49892 |
| $1s5s^1S$ | -2.02117316 | −2.0211762 | −2.021176531 | 0.49947116 | 0.499471[a] | 0.4993±0.0019 | 0.499465 |
| $1s6s^1S$ | -2.01455645 | −2.0145627 | | 0.49970073 | 0.499701[a] | | 0.499665 |
| | | | | | | | |
| $1s2s^3S$ | -2.17522937 | | −2.175229378 | 0.50037593 | 0.500376[e] | 0.50038±0.00015 | 0.500378 |
| $1s3s^3S$ | -2.06868906 | | −2.068689045 | 0.50007327 | | 0.50019±0.00024 | 0.5000736 |
| $1s4s^3S$ | -2.03651200 | | −2.036512038 | 0.50002655 | | 0.49993±0.00038 | 0.5000267 |
| $1s5s^3S$ | -2.02261852 | | −2.022618670 | 0.50001261 | | 0.50012±0.00048 | 0.5000127 |
| $1s6s^3S$ | -2.01537422 | | | 0.50000683 | | | 0.5000070 |

[a] Ref. [6]; [b] Ref. [8]; [c] Ref. [15]; [d] Ref. [22]; [e] Ref. [9]

Table IX. Energies and von Neumann entropies of the singlet and triplet $S$ states in the helium atom.

| State | Energy (a.u.) | $S_{vN}$ | |
|---|---|---|---|
| | Present work | Present work | Others |
| $1s^2$ $^1S$ | -2.90372437 | 0.08489987 | 0.0785[a]<br>0.085022[b]<br>0.06749889[c]<br>0.0675[d] |
| $1s2s$ $^1S$ | -2.14597404 | 0.99191721 | 0.991099[a]<br>0.991917[b] |
| $1s3s$ $^1S$ | -2.06127196 | 0.99873620 | 0.998513[a] |
| $1s4s$ $^1S$ | -2.03358497 | 0.99967147 | 0.999577[a] |
| $1s5s$ $^1S$ | -2.02261852 | 0.99990742 | 0.999838[a] |
| $1s6s$ $^1S$ | -2.01537422 | 0.99997755 | 0.999923[a] |
| | | | |
| $1s2s$ $^3S$ | -2.17522937 | 1.00552680 | 1.00494[a]<br>1.005527[b] |
| $1s3s$ $^3S$ | -2.06868906 | 1.00125237 | 1.00114[a] |
| $1s4s$ $^3S$ | -2.03651200 | 1.00049300 | 1.000453[a] |
| $1s5s$ $^3S$ | -2.02261852 | 1.00024725 | 1.000229[a] |
| $1s6s$ $^3S$ | -2.01537422 | 1.00014076 | 1.000133[a] |

[a] Benenti *et. al.* [7]; [b] Lin and Ho [9]; [c] Hofer [22]; [d] Huang *et. al.* [23]

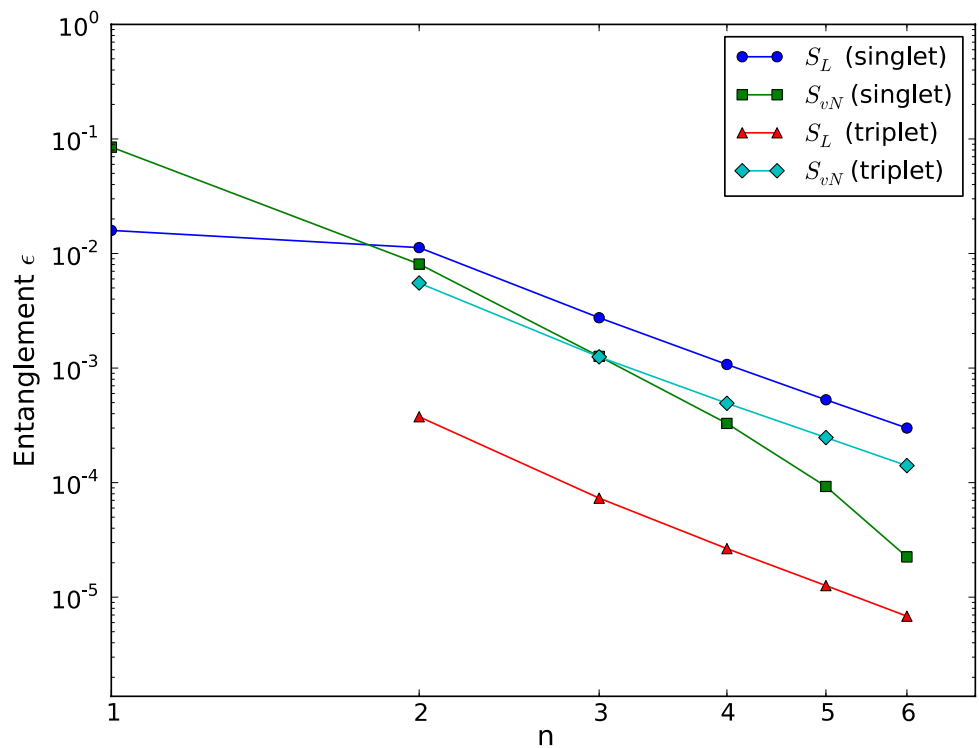

Figure I. The absolute entanglement entropies (see Eq. (18) for definition) for the $1s^2\ ^1S$ to $1sns\ ^1S$ states and $1s2s\ ^3S$ to $1sns\ ^3S$ states, up to $n=6$, in log-log plot.

# References

[1] Manzano, D., Plastino, A. R., Dehesa J. S. and Koga, T.: Quantum entanglement in two-electron atomic models . *Journal of Physics A: Mathematical and Theoretical* **43**, 275301 (2010).

[2] Coe, J. P. and D'Amico, I.: The entanglement of few-particle systems when using the local-density approximation. *Journal of Physics: Conference Series* **254**, 012010 (2010).

[3] Dehesa, J. S., Koga, T. Yanez, R. J. Plastino, A. R. and Esquivel, R. O.: Quantum entanglement in helium . *Journal of Physics B: Atomic, Molecular and Optical Physics* **45**, 015504 (2012).

[4] Dehesa, J. S., Koga, T. Yanez, R. J. Plastino, A. R. and Esquivel, R. O.: Corrigendum: Quantum entanglement in helium. *Journal of Physics B: Atomic, Molecular and Optical Physics* **45**, 239501 (2012).

[5] Osenda, O. and Serra, P.: Scaling of the von Neumann entropy in a two-electron system near the ionization threshold. *Phys. Rev. A* **75**, 042331 (2007).

[6] Lin, Y.-C., Lin, C.-Y. and Ho, Y. K.: Spatial entanglement in two-electron atomic systems. *Phys. Rev. A* **87**, 022316 (2013).

[7] Benenti, G., Siccardi, S. and Strini, G., Entanglement in helium. *Euro. Phys. J. D* **67**, 83 (2013).

[8] Lin, C.-H., Lin, Y.-C., Ho, Y. K., Quantification of Linear Entropy for Quantum Entanglement in He, H− and Ps− Ions Using Highly-Correlated Hylleraas Functions, *Few-Body Systems.* **54**, 2147 (2013).

[9] Lin, Y.-C. and Ho, Y. K., Quantum entanglement for two electrons in the excited states of helium-like systems, e-print: arXiv:1307.5532

[10] Abdullah, S., Coe, J. P. and D'Amico, I.: Effect of confinement potential geometry entanglement in quantum dot-based nanostructures. *Phys. Rev. B* **80**, 235302 (2009).

[11] Nazmitdinov, R. G., Simonovi ́c, N. S., Plastino, A. R. and Chizhov, A. V.: Shape transitions in excited states of two-electron quantum dots in a magnetic field . *J. Phys. B: At. Mol. Opt. Phys.* **45**, 205503 (2012).

[12] Okopinska, A. and Koscik P.: Correlation and Entanglement in Elliptically Deformed Two-Electron Quantum Dots. *Few-Body Syst*. **50**, 413 (2011).

[13] Coden, D. S. A., Romero, R. H., Ferr ́on, A. and Gomez, S. S.: Impurity effects in two-electron coupled quantum dots: entanglement modulation . *J. Phys. B: At. Mol. Opt. Phys.* **46**, 065501 (2013).

[14] Schroter, S., Friedrich, H, and Madronero, J.: Considerations on Hund's first rule in a planar two-electron quantum dot. *Phy. Rev. A* **87**, 042507 (2013).

[15] Kościk, P., Entanglement in S states of two-electron quantum dots with Coulomb impurities at the center, *Phys. Lett. A* **377**, 2393 (2013).

[16] Koscik, P.: Quantum Correlations of a Few Bosons within a Harmonic Trap , *FEW-BODY SYSTEMS* **52**, 49 (2012)

[17] Okopinska, A; Koscik, P.: Entanglement of Two Charged Bosons in Strongly Anisotropic Traps. *FEW-BODY SYSTEMS* **54**, 629 (2013).

[18] Naik, D. S. Peterson, C. G. White, A. G. Berglund, A. J. and Kwiat, P. G.: Entangled state quantum cryptography: Eavesdropping on the Ekert protocol. *Phys. Rev. Lett.* **84**, 4733 (2000).

[19] Zhao, M.-J., Fei, S.-M. and Li-Jost, X.: Complete entanglement witness for quantum teleportation. *Phys. Rev. A* **85**, 054301 (2012).

[20] Nielsen, M. A. and Chuang, I. L.: *Quantum Computation and Quantum Information* (Cambridge University Press, Cambridge, 2000).

[21] Tichy, M. C., Mintert, F. and Buchleitner, A.: Essential entanglement for atomic and molecular physics, *J. Phys. B: At. Mol. Opt. Phys.* **44**, 192001 (2011).

[22] Hofer, T. S.: On the basis set convergence of electron-electron entanglement measures: helium-like systems, *Front. Chem.: Theor. Comput. Chem.* **1**, 24 (2013).

[23] Huang, Z., Wang, H. and Kais, S.: Entanglement and electron correlation in quantum chemistry calculations, *Journal of Modern Optics* **53**, 2543 (2006).

[24] Tan, S. S. and Ho, Y. K.: Determination of Resonance Energy and Width by Calculation of the Density of Resonance States Using the Stabilization Method, *Chin. J. Phys.* **35**, 701 (1997).

[25] Ho, Y. K.: Recent Advances in the Theoretical Methods and Computational Schemes for Investigations of Resonances in Few-Body Atomic Systems, *Few-Body Syst.* **54,** 31 (2013).